\documentclass[preprint]{aastex63}
\usepackage{amsmath}
\usepackage{amssymb}
\usepackage{graphicx}
\usepackage{color}
\usepackage{xcolor}
\usepackage{xfrac}
\usepackage[caption=false]{subfig}
\usepackage{rotating}
\usepackage{hyperref}
\usepackage[utf8]{inputenc}

\begin{document} 

\title{Axial Asymmetry Studies 
in {\it Gaia} 
Data Release 2 \\ Yield the Pattern Speed of the Galactic Bar}

\author[0000-0002-9785-914X]{Austin Hinkel}
\affiliation{Department of Physics and Astronomy, 
University of Kentucky, Lexington, KY 40506-0055}
\author[0000-0002-6166-5546]{Susan Gardner}
\affiliation{Department of Physics and Astronomy, 
University of Kentucky, Lexington, KY 40506-0055}
\author[0000-0002-9541-2678]{Brian Yanny}
\affiliation{Fermi National Accelerator Laboratory, Batavia, IL 60510}

\date{\today}

\begin{abstract}
Our recent studies of axial-symmetry breaking 
in the nearby ($d <3 \,{\rm kpc}$) star counts are sensitive
to the distortions of stellar orbits 
perpendicular and parallel to the orientation of the bar just 
within and beyond the 
outer Lindblad resonance (OLR) 
radius.
Using 
the location of
the sign flip in the left-right asymmetry  in 
stars counts about the anticenter 
line 
to determine the OLR radius $R_{\rm OLR}$, 
and treating the bar as if it were a weakly non-axisymmetric 
effect, 
we use
$R_{\rm OLR}$ and recent measurements of the Galactic rotation 
curve  
and the Sun--Galactic-center
distance $R_{0}$ 
to determine the pattern speed $\Omega_{\rm p}$ of the Galactic bar, 
as well as the Galactic corotation radius $R_{\rm CR}$. After removing the effect of
the Large and Small Magellanic clouds 
from our asymmetry measurement, 
we find that $R_{\rm OLR}=(0.96 \pm 0.03)R_0 = 7.85 \pm 0.25 \ \rm kpc$, $\Omega_{\rm p} = 49.3 \pm 2.2 \ \rm km \  s^{-1} \ kpc^{-1}$, $R_{\rm CR}=(0.58 \pm 0.04)R_0 = 4.76 \pm 0.27 \ \rm kpc$, revealing, as we shall show, that the Milky Way's bar is 
likely both weak and fast, though we also note possible evidence for non-steady-state effects in 
the bar region.

\end{abstract}

\section{Introduction} \label{sec:Intro}

It is well established that there is a bar at the center of the Galaxy \citep{gerhard2015galactic} and that this structure 
rotates in a manner such that its stars and dust have net
motion in the bar rest frame~\citep{binney2008GD}. 
The pattern speed, $\Omega_{\rm p}$, is the assessment of this rotation of the bar's potential, and 
models of that unknown potential are ordinarily 
needed in order to explain the motion of certain stellar populations to infer properties of the bar. 
This theoretical barrier, along 
with 
observational issues 
associated with 
high source densities, extinction, and reddening
in the central region of the Galaxy, have resulted in 
a wide array of values, differing by more than a factor of two,
for $\Omega_{\rm p}$
\citep{bland-hawthorne2016galaxy}.
To illustrate, various methods 
\citep{dehnen2000effect,debattista2002pattern,chakrabarty2007phase,minchev2007new,antoja2014constraints} favor a fast bar, such as
the pattern speed of $\Omega_{\rm p} = 57.4^{+2.8}_{-3.3} \ \rm km \ s^{-1} \ kpc^{-1}$ \citep{chakrabarty2007phase}, whereas studies in the
Galactic bar region \citep{portail2015mass, portail2016dynamical,sanders2019pattern,bovy2019life}
can find considerably slower values, 
such as 
$\Omega_{\rm p} = 25 - 30 \ \rm km \ s^{-1} \ kpc^{-1}$ \citep{portail2015mass}.  
Bearing in mind the varied pictures and mechanisms employed in determining the pattern speed, the review of 
\citet{bland-hawthorne2016galaxy} 
give a recommended range of 
$\Omega_{\rm p} = 43 \pm 9 \ \rm km \ s^{-1} \ kpc^{-1}$.
A model-independent method of 
measuring the pattern speed that utilizes the continuity equation does exist, however, if the pattern is steady
\citep{tremaine1984kinematic, debattista2002pattern, sanders2019pattern}
but implementing
it requires 
proper motion 
information for stars in the
Galactic Bar. Recently \citet{sanders2019pattern} have used 
{\it Gaia} Data Release 2 (DR2) 
and VISTA Variables in the Via Lactea (VVV)
data \citep{Minniti2010} to find 
$\Omega_{\rm p} = 41 \pm 3 \ \rm km \ s^{-1} \ kpc^{-1}$, 
where the error is statistical only, with an additional 
suggested systematic uncertainty of 
$5-10 \ \rm km \ s^{-1} \ kpc^{-1}$. 

The wide range of reported 
pattern speeds is also partly responsible for the 
wide range of radii associated with resonant effects 
driven by the Galactic bar: that is, 
the radius of the Outer Lindblad resonance (OLR) and the radius of the corotation resonance (CR). As such, it is unclear
whether the stellar streams seen in the solar vicinity \citep{raboud1998evidence, dehnen1999pattern, fux2001order, sellwood2010recent}
are due to 
a CR \citep[e.g.][]{mishurov1999yes} or an OLR \citep[e.g.][]{dehnen2000effect} or a $4:1$ 
OLR 
\citep{hunt2018outer}. Until recently \citep{HGY20}, there has been 
no 
model-independent 
way of discriminating between the 
possibilities in the existing data.

This lack of consensus regarding
the pattern speed may come, in part, from the use of astrometric/photometric methods \citep{debattista2002pattern,sanders2019pattern,bovy2019life}
or of dynamical methods \citep{englmaier1999gas,portail2015mass, portail2016dynamical}, and
this 
spills over into the debate on the location of the resonances of the Galactic bar.  
Moreover, it has been 
suggested that the inconsistencies between
the two sorts of methods can be reduced
by having the bar rotate at a slower speed today than it has in the past \citep{monari2017tracing}.
The findings of \citet{sanders2019pattern} may yield a simpler explanation:  
systematic effects from dust, e.g., 
tend to 
lower assessments of the pattern speed artificially, 
especially when observations of stars from the far side of the galactic center are used. 
Namely, 
\citet{sanders2019pattern} find 
$\Omega_{\rm p} = 41 \pm 3 \ \rm km \ s^{-1} \ kpc^{-1}$ 
and 
$\Omega_{\rm p} = 31 \pm 1 \ \rm km \ s^{-1} \ kpc^{-1}$ for 
stars in the near side of the bar and in both the near and far sides, respectively,  providing the basis for their systematic error assessment.  Alternatively, \citet{hilmi2020fluctuations}
suggest that the bar's length and pattern speed can fluctuate by as much as 20\% as the bar interacts with nearby spiral arms, 
perhaps explaining 
the different estimates of $\Omega_{\rm p}$ from different methods.  
\citet{hilmi2020fluctuations} note that the pattern speed as inferred from outer disk dynamics should 
reveal the time-averaged value of $\Omega_{\rm p}$, as opposed to instantaneous values 
measured in the central region via astrometric or photometric methods.

For a given galactic rotation curve, the pattern speed 
sets 
where these resonances are located. Thus, the determination
of a resonant radius can also be used to fix the pattern 
speed, with information on additional resonant radii 
giving further information on the morphology of the bar. 
As motivated by leading order perturbation theory in 
the strength of the nonaxisymmetric bar 
potential \citep{binney2008GD},
stars in resonant orbits between the 
Inner Lindblad resonance (ILR) and the CR are oriented along the bar, stars between the CR and the OLR orbit with trajectories perpendicular to the bar, and beyond the OLR the stellar orbits tend to be elongated along the bar's orientation \citep{contopoulos1980orbits}. These features are expected
to persist even as the bar potential grows strong, though
the fractional number 
of stars following the particular orbits predicted
by leading-order perturbation theory may grow small \citep{binney2008GD}. 
Nevertheless, by using the change in sign of the axial asymmetry in star counts \citep{GHY20, HGY20} to determine the location 
of the OLR and using leading order perturbation theory 
to determine the pattern speed as well as the CR, we find 
that our determined CR is crudely commensurate 
with the length of the Galactic bar 
 --- this is expected if the 
Galaxy's bar is indeed weak \citep{aguerri1998bar}.

In this letter, we employ a novel, 
model-independent method for determining the bar's pattern speed and resonant effects by leveraging 
our ability to detect axially asymmetric orbits.  
From tests of axisymmetry of our galaxy \citep{GHY20}, \citet{HGY20} determine the radius of the OLR using {\it Gaia} DR2 data \citep{prusti2016gaia, brown2018gaia, lindegren2018gaia}, 
and here we use this measurement along with 
leading order perturbation theory \citet{binney2008GD} and the rotation curve of \citet{eilers2019circular} in order to obtain a measurement of the pattern speed.  
With this we can also determine the 
radius of the CR.\footnote{
Our analysis uses the rotation curve of \citet{eilers2019circular}, which assumes 
$R_{0} = 8.122(31)$ kpc \citep{abuter2018detection}, whereas we employ a subsequent (and more precise) 
determination of the Sun--Galactic-center distance, 
$R_{0} = 8.178(26)$ kpc \citep{abuter2019geometric} 
as appropriate.}
We also document an abrupt change in the vertical structure of the galaxy very near to the OLR; we believe this speaks
to north-south differences in the Galactic bar or perhaps some interaction between the OLR and separate north-south differences in the plane \citep{widrow2012galactoseismology, yanny2013stellar, ferguson2017milky, bennett2018vertical}. 
We 
note, for reference, that 
a significant north-south asymmetry has been 
recently suggested in the 
galactic center excess 
\citep{leane2020enigmatic}.
Finally, we compare our results 
with those already in the literature, as well as with other established features of the bar, noting the additional possibility of non-steady-state and/or axial-symmetry-breaking effects 
in the bar region.

\section{Theory} \label{sec:Theory}

As motivated through the perturbation theory analysis of 
\citet{binney2008GD} and depicted graphically in \citet{dehnen2000effect}, the Galactic bar drives the OLR,
holding 
sway over the shape of stellar orbits despite the affected stars not being within the 
physical extent of the bar, at Galactocentric, in-plane 
$R < {\ell}_{\rm bar}$, where ${\ell}_{\rm bar}$ is the bar half-length.  Due to the periodic nature of the bar's gravitational force on stars 
at 
$R > {\ell}_{\rm bar}$, stars may receive a pull from the bar at the same phase in their orbit, exciting the orbit into an elliptical shape.  For stars just inside (outside) the radius of the OLR, orbits are elongated perpendicular (parallel) to the bar \citep{dehnen2000effect, contopoulos1980orbits}, 
which has been thought to point  
at $\sim 10^\circ-70^\circ$ \citep{dehnen2000effect}
away from the Sun-Galactic center line 
($\phi=180^{\circ}$), with 
more recent work \citep{robin2012stellar,portail2016structure, anders2019photo} finding values
within $13^\circ - \sim 40^{\circ}$.

Given that this effect has 
$\phi$-dependence, it 
breaks axial symmetry and thus can result in a 
measurably non-zero value of the axial asymmetry, ${\cal A}$, about the anti-center line as defined in \citet{GHY20}.  Indeed, one would expect that the stars ``promoted'' to higher $R$ by 
the bar near the OLR would cause a very slight over-density over a small range in azimuth near the bar's principal axis at some value $R_{\rm OLR} + \Delta R$ and leave behind a commensurate, slight under-density at some $R_{\rm OLR} - \Delta R$.  By scanning over various values of $R$
we have found 
that ${\cal A}$ varies radially \citep{HGY20}.

\begin{figure}
    \centering
    \includegraphics[scale=0.55]{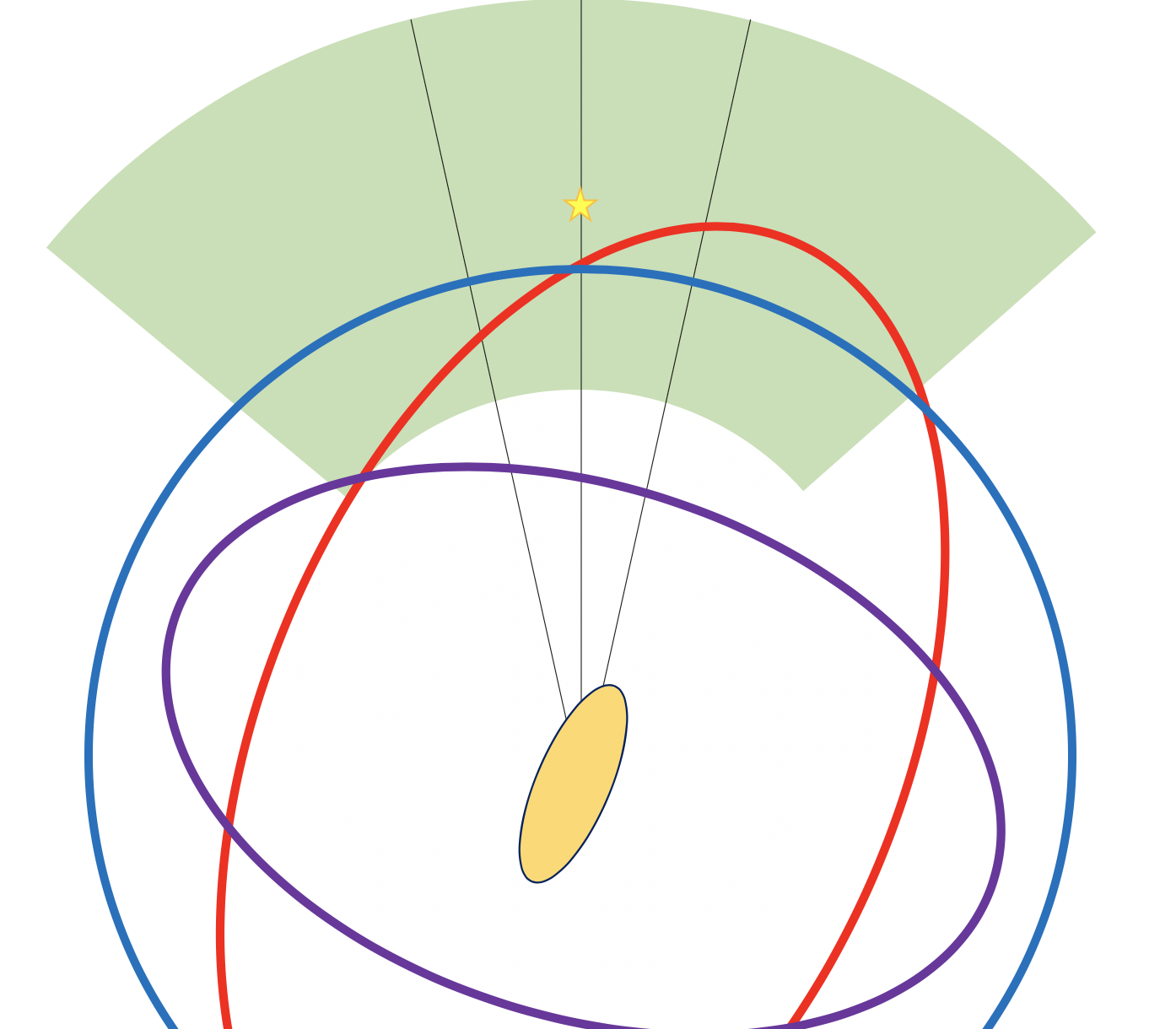}
    \caption{
    A schematic depiction of the orbital alignments due to 
    the bar in the OLR region.  The blue circle is the orbital radius of the OLR, the purple  ellipse is an orbit interior to the OLR, and the red ellipse is an orbit exterior to the OLR.  The green annular wedge region is our sample's in-plane footprint, with a star signifying the sun's position, and  the yellow ellipse is the Galactic bar.  Stellar orbits tend to align parallel (perpendicular) to the bar when the orbit is just outside (inside) the Outer Lindblad resonant radius.  
    The geometry has been greatly exaggerated
    and we have shown closed orbits only,
    in order to 
    illustrate 
    the small effect we have found}. 
    \label{fig:olr_schematic}
\end{figure}

The orbital alignments due to the central bar in the OLR region 
break axial asymmetry in the manner illustrated schematically in Fig.~\ref{fig:olr_schematic}.  Just outside the resonant radius, we expect to find more stars to the right of the $\phi = 180^{\circ}$ line ($\phi < 180^{\circ}$), and expect to find more stars on the left ($\phi > 180^{\circ}$) when just inside the resonant radius.  Thus, as one moves outward in 
$R$ the expected axial asymmetry would go from left-heavy to right-heavy, corresponding to a sign flip:
\begin{equation}
    {\cal A}(R < R_{\rm OLR}) > 0 \ \longrightarrow \ 
    {\cal A}(R > R_{\rm OLR}) < 0.
\end{equation}
As such, the value of $R$ that yields zero asymmetry is the location of the sign-flip and thus the location of the OLR.
In contrast, if the axially asymmetric effect were, 
rather, a CR, then the sense of the sign flip would 
change from ${\cal A} < 0 \, \longrightarrow {\cal A} >0$ 
as $R$ increases.

Following the methods of \citet{binney2008GD}, 
a non-axisymmetric contribution to the Galactic 
gravitational potential 
can be treated as a weak perturbation. 
Working in 
a 
reference frame 
rotating with the bar, at a steady pattern speed 
$\Omega_{\rm p}$, 
we have the Lagrangian
\begin{equation}
    L = \frac{1}{2} \dot R^2 + \frac{1}{2}[R(\dot \varphi + \Omega_{\rm p})]^2 - \Phi(R, \varphi) \,,
    \label{lagrangian}
\end{equation}
where we employ cylindrical coordinates with $\varphi=0$ 
aligned along its long axis.
The potential can be broken into an unperturbed, axisymmetric potential and a 
non-axisymmetric correction:
\begin{equation}
    \Phi(R, \varphi) = \Phi_{\rm u}(R) +  \Phi_{1}(R, \varphi) \,.
    \label{potl}
\end{equation}
In the absence of the perturbation, we find a 
circular orbit at $R$ with 
$\dot\varphi = \Omega - \Omega_{\rm p}$, where 
the frequency
\begin{equation} 
\Omega = \pm \sqrt{\frac{1}{R} \frac{d\Phi_u }{dR}} \,,
\end{equation}
where $\Omega >0$ corresponds to prograde rotation. 
Specifying 
the form of the perturbing potential as per 
\citep{binney2008GD} we have 
\begin{equation}
    \Phi_1(R, \varphi) = \Phi_{\rm bar}(R) {\rm cos}(m \varphi) 
    \,,
\end{equation}
where $m=2$ for a Lindblad resonance. 
Now with $R(t) = R_{\rm u} + R_1(t)$ and
$\varphi(t) = \varphi_{\rm u} (t) + \varphi_{1} (t)$, 
analyzing
the equations of motion while 
working to leading order in 
$|\Phi_1/\Phi_{\rm u}| \ll 1$ 
and assuming $\varphi_{1} \ll \varphi_{\rm u}$ 
yields 
\begin{equation}
    \ddot R_1 + \kappa_{0}^2 R_1 = - \left[ \frac{d \Phi_{\rm bar}}{dR} 
    + \frac{2 \Omega \Phi_{\rm bar}}{R(\Omega-\Omega_{\rm p})} \right]_{R = R_{\rm u}} {\rm cos}(m(\Omega(R_{\rm u}) -
    \Omega_{\rm p})t) \,.
\end{equation}
where 
$\kappa_{0}$ is the natural harmonic frequency for the oscillatory perturbation provided by the bar, 
\begin{equation}
    \kappa_{0}^{2} = 
    \left( \frac{d^2 \Phi_{\rm u}}{dR^2} 
    + 3 \Omega^2 \right)_{R = R_{\rm u}}
    = \left( R \frac{d \Omega^2}{dR} 
    + 4 \Omega^2 \right)_{R = R_{\rm u}}  
    \,,
    \label{kappaDefinition}
\end{equation}
and the general solution 
\begin{equation}
R_1(t) = A \cos (\kappa_0 t + \alpha) 
- \left[ \frac{d \Phi_{\rm bar}}{dR} 
    + \frac{2 \Omega \Phi_{\rm bar}}{R(\Omega-\Omega_{\rm p})} \right]_{R = R_{\rm u}} 
\left(    \frac{{\rm cos} (m(\Omega - \Omega_{\rm p})t)}
    {\kappa_0^2 - m^2 (\Omega - \Omega_{\rm p})^2} \right) \,,
\label{r1eq}
\end{equation}
so that open orbits appear with nonzero, arbitrary  
$A$ for any $\alpha$.  Regardless, 
a resonance appears if 
$\kappa_0^2 - m^2 (\Omega - \Omega_{\rm p})^2 = 0$, and it is 
an $m=2$ OLR 
if 
\begin{equation}
    \Omega_{\rm p} - \Omega = \kappa_{0} / 2 \,.
    \label{OuterLindbladCondition}
\end{equation}
Notice this condition can be 
combined with Eq.~\ref{kappaDefinition} 
to yield:
\begin{equation}
    4 (\Omega_{\rm p} - \Omega)^{2} \bigg\rvert_{R = R_{\rm OLR}} = 
    \left( R\frac{d \Omega^2}{dR} 
    + 4 \Omega^2 \right)_{R = R_{\rm OLR}} \,,
    \label{FindPatternSpeedSetUp}
\end{equation} 
to give the pattern speed 
from $R_{\rm OLR}$ and the $R$-dependence of $\Omega$:
\begin{equation}
    \Omega_{\rm p} = \Omega(R_{\rm OLR}) + \frac{1}{2} \sqrt{4\Omega^2(R_{\rm OLR}) + R_{\rm OLR}\frac{d \Omega^2}{dR}\bigg\rvert_{R = R_{\rm OLR}}} \,.
    \label{PatternSpeed}
\end{equation} 
Finally, the pattern speed determines the CR radius:
\begin{equation}
    \Omega_{\rm p} = \Omega(R_{\rm CR}) \,,
    \label{CRcondition}
\end{equation}
noting that we cannot also determine the location of the 
Inner Lindblad resonance (ILR) with these methods for 
want of information on $\Omega$ with $R$ in 
the very inner portion of our galaxy.

To determine the numerical 
value of the pattern speed and more, we 
use an observational assessment of the 
Galactic rotation curve, which yields both 
$\Omega^2$ and $d\Omega^2/dR$ with $R$.
That is, the Galactic 
rotation curve is the circular speed $v_{\rm c}$ with $R$, where 
\begin{equation}
    \Omega \equiv \frac{v_{\rm c}(R)}{R} = 
    \sqrt{\frac{1}{R}\frac{d\Phi_{\rm u}}{dR}}\,.
    \label{OmegaCurve}
\end{equation}
For this, we use the recent, high precision determination 
of \citet{eilers2019circular}, which uses an analysis of 
red-giant branch stars from {\it Gaia} DR2, cross-matched
with APOGEE data, for refined distance assessments
\citep{hogg2019spectrophotometric}.
The analysis itself uses a Jeans equation framework in 
which the underlying Galactic distribution function 
$f({\bf{ x}},{\bf{ v}},t)$
is assumed to be axially symmetric and in steady state. 
This yields 
\begin{equation}
v_{\rm c}^2 = \langle v_\phi \rangle^2 - 
\langle v_R \rangle^2\left( 1 + 
\frac{\partial \rm{ln} \langle v_R^2 \rangle }{\partial {\rm ln} R} + 
 \frac{\partial \rm{ln} \nu }{\partial {\rm ln} R}
\right) + \delta 
\end{equation}
where 
$\nu ({\bf{x}},t) 
= \int d^3{{\bf v}} f({\bf{x}}, {\bf{v}}, t)$ 
and $\delta=0$. We can, however, determine the 
modification of $v_{\rm c}^2$ were all the neglected terms
included. This gives 
\begin{equation}
\delta = 
-R \frac{\partial }{\partial t}
\left( \langle v_R \rangle {\rm ln}\nu\right)
-\frac{\partial }{\partial \phi}
\left( \langle v_R v_\phi \rangle {\rm ln}\nu\right)
- R \frac{\partial }{\partial z}
\left( \langle v_R v_z \rangle {\rm ln}\nu\right) \,,
    \end{equation}
    where the additions reflect corrections for 
    non-steady-state, axial-symmetry-breaking, 
    and $z$-dependent effects, respectively. 
The $z$-dependent term also appears in 
\citet{eilers2019circular} and is estimated to affect
$v_{\rm c}$ at the $\sim$1\% level at $R\sim 18\,{\rm kpc}$.
The axial symmetry breaking term vanishes if 
$\nu({\bf{x}})$ itself is axially symmetric. 
We will note a possible role for these small terms,
likely characterized in size by the non-steady-state term, 
later. 
\citet{eilers2019circular} determines $v_{\rm c}(R)$ over
$5 \stackrel{<}{{}_{\sim}} R 
\stackrel{<}{{}_{\sim}} 25$ kpc, 
for which they report 
the linear 
parametrization 
\begin{equation}
v_{\rm c} (R) = (229.0 \pm 0.2) \rm{km} \, {\rm s}^{-1} - 
(1.7 \pm 0.1) \rm{km \, s}^{-1} \, \rm{kpc}^{-1} \cdot 
(R- R_0) \,,
\label{vcparam}
\end{equation}
where here 
$R_0 = 8.122(31)$ kpc \citep{abuter2018detection} has
been employed. We employ this parametrization in what 
follows.

\section{Analysis} \label{sec:Analysis} 

As we showed in \citet{GHY20}, effects from the LMC and Galactic Bar are the two dominant contributors of axial symmetry breaking in the solar neighborhood.  Further, in \citet{HGY20}, we found a sign flip in the sense of the asymmetry that matches 
that expected from an OLR assuming the 
determined bar orientation \citep{robin2012stellar,portail2016structure,anders2019photo} 
does indeed point in the third quadrant of the galactocentric 
rectangular coordinate system in which the positive 
$x$-axis 
points from the GC in the direction opposite the sun 
with $y$ and $z$ following from 
a right-hand coordinate system choice in which $z$ increases from zero at the mid-plane to larger values toward the North Galactic Pole.  Here, we refine the sign flip analysis in order to remove any background effects from the overall distortion of the galaxy due to the LMC's influence, 
which we found to be described by a prolate shape pointing
towards the LMC \citep{GHY20, erkal2019total}. 
We expect this global background effect to be a constant offset over the volume of space we study, and we define this background asymmetry as $<{\cal A}>_{\rm B}$.  As such, the precise value of $R$ where the equality  
\begin{equation}
    <{\cal A}(R)> - <{\cal A}>_{\rm B} = 0
    \label{backgroundSubtract}
\end{equation}
corresponds to the radius of the OLR.
We estimate the background asymmetry by integrating over the entire volume of the sample of \citet{GHY20} and find that $<{\cal A}>_{\rm B} = -0.0032 \pm 0.0003$. This moves our measurement of the sign flip from \citet{HGY20}, and thus $R_{\rm OLR}$, slightly outward in $R$, as expected.  

\begin{figure}[ht!]
\begin{center}
\subfloat[]{\includegraphics[scale=0.55]{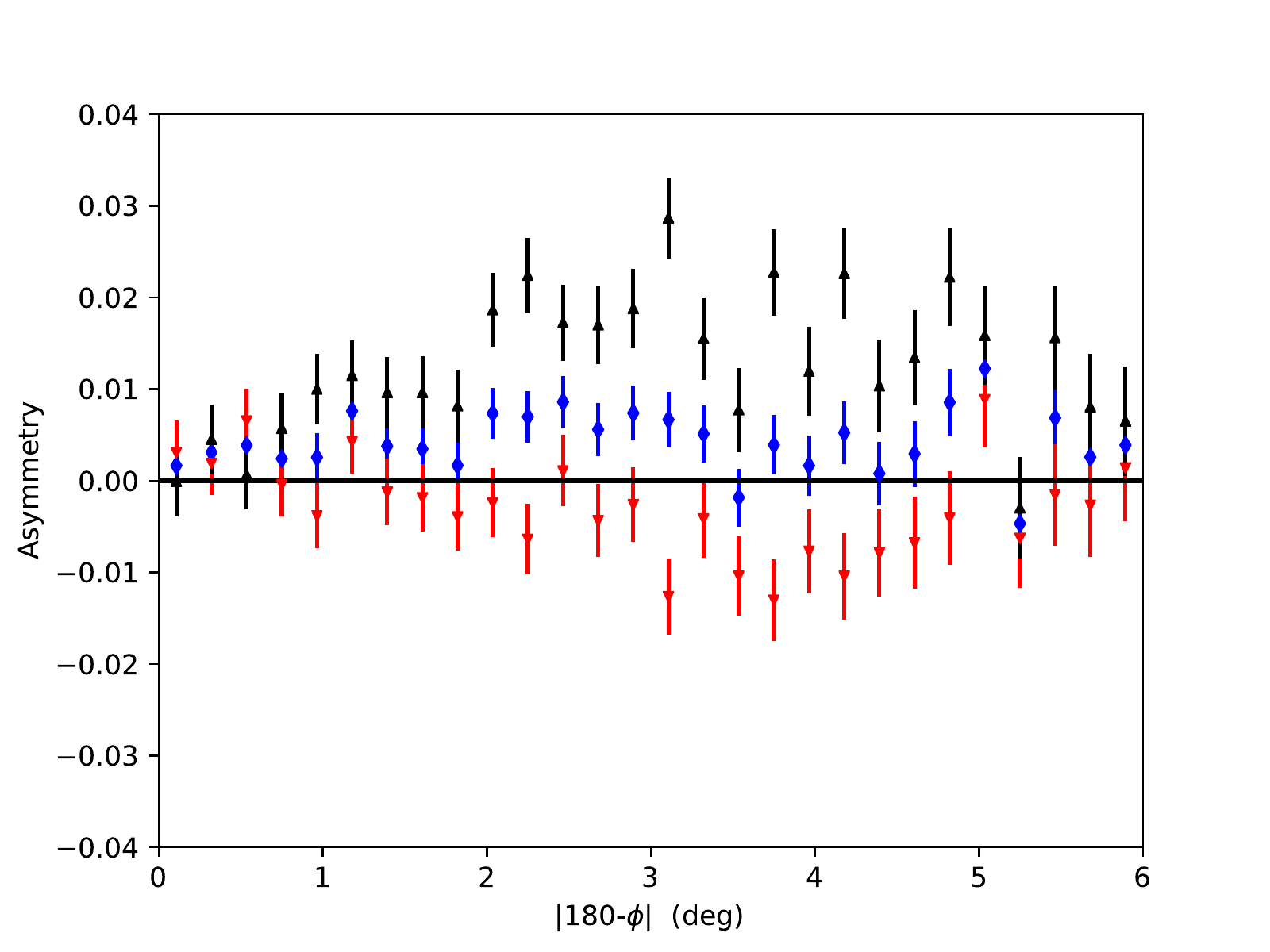}}
\subfloat[]{\includegraphics[scale=0.55]{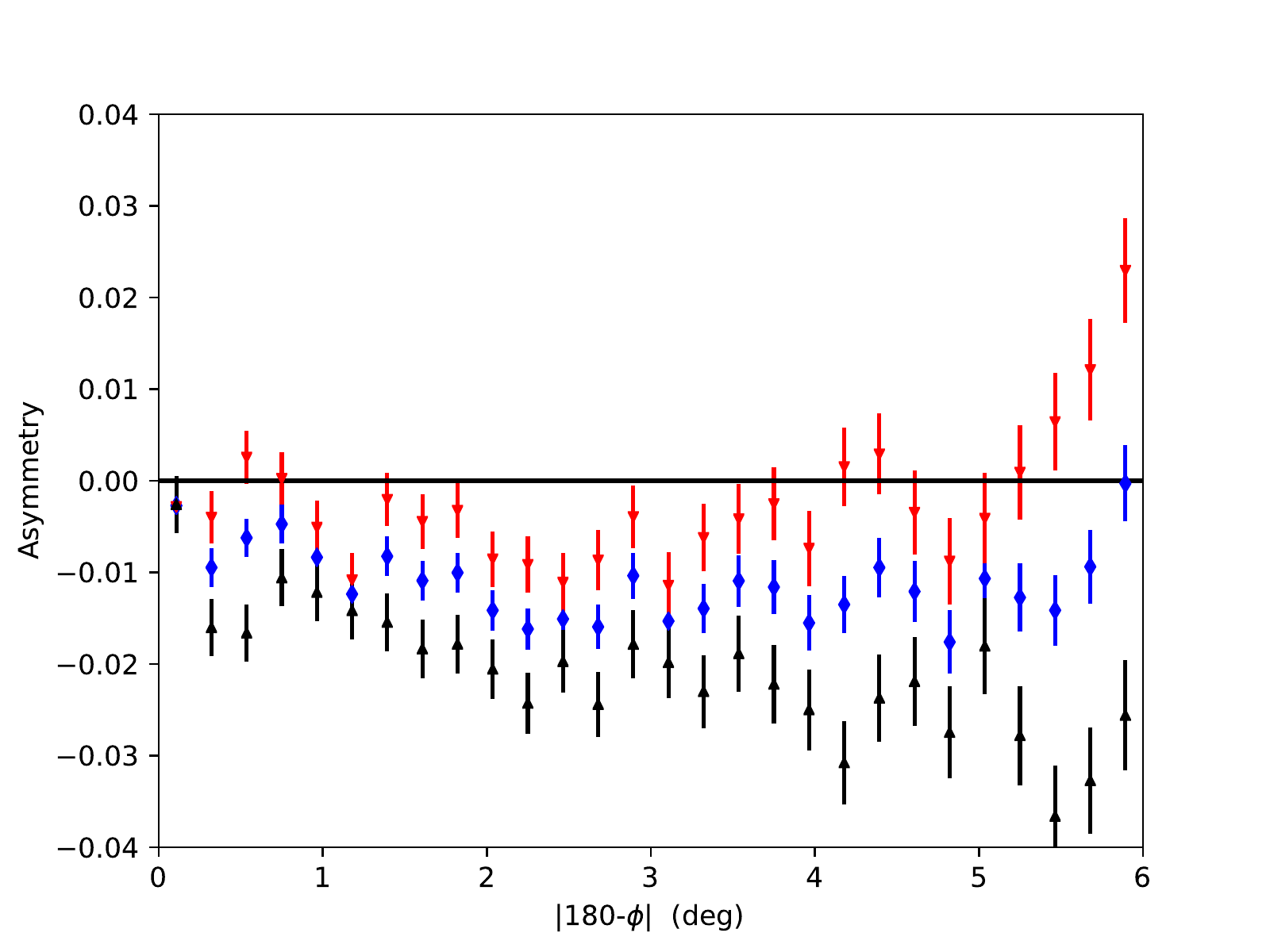}}
\caption{
(a) The axial asymmetry for $R \in [0.8975 , 0.9600]R_0$.
(b) The axial asymmetry for $R \in [0.9600 , 1.0225]R_0$.
The blue diamonds are the aggregate axial asymmetry and the black and red triangles are for the northern ($z > 0$) and southern ($z < 0$) halves respectively.  The sign flip in the aggregate asymmetry is clearly visible here, which we attribute to the bar's OLR.  In addition, the vertical structure changes just beyond the OLR, with a 
north-left
correlation for $R < R_{\rm OLR}$ and a 
north-right correlation for $R > R_{\rm OLR}$.
}
\label{fig:verticalChange}
\end{center}
\end{figure}

In practice, we 
repeat the radial scans of \citet{HGY20} and subtract the offset in order to find the bin with zero asymmetry. 
The results of this analysis are tabulated in Table~\ref{tab:ROLR}.
The resulting shift in the determined OLR location appears in 
Table~\ref{tab:patternspeedvals}. 
The OLR radius is defined as the center of the bin in Table~\ref{tab:ROLR} which, after accounting for the background asymmetry, yields an asymmetry that is within 1-$\sigma$ from 0.  Note that, after rounding, this yields $R_{\rm OLR} = (0.96 \pm 0.03)R_0 = 7.85 \pm 0.25 \ \rm kpc$ where the uncertainty in the OLR radius assessment is the first $\Delta R$ in the successively smaller $\Delta R$ scans in which a ``zero'' is no longer discernible in a single bin, rounded to one significant figure.
The measured 
axial asymmetry just within and beyond the determined OLR 
location in $R$ is shown 
in Fig.~\ref{fig:verticalChange}. 
We 
discuss its interesting north/south differences in the next section. Here we wish to focus on 
the size of the asymmetry 
 $<{\cal A}(R)> - <{\cal A}>_{\rm B}$ 
itself because this is reflective of the number of stars
that populate the distorted orbits we have analyzed. As tabulated in Table~\ref{tab:DeltaRanalysis}, the flip in sign of the asymmetry is quite 
symmetric about the Outer Lindblad resonant radius, which is  expected if the stars are excited to higher $R$ and leave behind a dearth of stars at lower $R$. Additionally, Table~\ref{tab:DeltaRanalysis} suggests that ${\cal O}(10^4)$ stars populate the distorted orbits that we analyze, corresponding to a small but statistically significant change in the sign of the asymmetry.

\begin{center}
\begin{table}[ht!]
    \centering
    \begin{tabular}{ |c|c|c|c|c| } 
       \hline
       $R_i - R_f$ ($R_{0}$)& $\Delta R$ ($R_{0}$)& $\langle {\cal A}(\phi) \rangle$ - $\langle {\cal A} \rangle_{\rm B}$ & $\sigma_{\langle \cal A \rangle}$ & Sign\\ 
       \hline
        0.8750 - 0.9375 & 0.0625 & +0.0103 & 0.0015 & +\\ 
       \hline
        0.9000 - 0.9625 & 0.0625 & +0.0067 & 0.0014 & +\\ 
       \hline
        0.9250 - 0.9875 & 0.0625 & +0.0005 & 0.0014 & 0\\ 
       \hline
       \hline
        0.9250 - 0.9625 & 0.0375 & +0.0049 & 0.0015 & +\\ 
       \hline
        0.9375 - 0.9750 & 0.0375 & +0.0009 & 0.0015 & 0\\ 
       \hline
        0.9500 - 0.9875 & 0.0375 & -0.0031 & 0.0014 & -\\ 
       \hline
       \hline
        0.9375 - 0.9625 & 0.0250 & +0.0037 & 0.0016 & +\\ 
       \hline
        0.9438 - 0.9688 & 0.0250 & +0.0013 & 0.0015 & 0\\ 
       \hline
        0.9500 - 0.9750 & 0.0250 & -0.0015 & 0.0015 & 0\\ 
       \hline
   \end{tabular}
       \caption{\label{tab:ROLR}
    { 
    Axial asymmetries, N+S, averaged over azimuthal angles about the
    anti-center direction up to $|180^{\circ}-\phi|=6^{\circ}$, 
    computed for a wedge of size $\Delta R$ for different choices of starting radius $R_i$, 
    with $R_f=R_i + \Delta R$, to reveal the sign change in the average asymmetry as 
    $R_i -R_f$ changes. We refine the location of the sign flip iteratively by computing the 
    average asymmetry with $R_i$ for smaller $\Delta R$. 
    Note that the distances are 
    in units of $R_0$ and
    that the ``Sign'' is assessed by whether the 
    magnitude 
    of the asymmetry difference is in excess of its error. 
    The uncertainty in the final asymmetry $\sigma_{\langle {\cal A} \rangle}$
    has been computed by adding the systematic axial asymmetry of \citet{HGY20} and statistical errors in quadrature and then adding the uncertainty from the background subtraction.
    }}
\end{table}
\end{center}

\begin{table}[]
    \centering
    \begin{tabular}{|c|c|c|c|c|}
        \hline
        $\Delta R \, (R_0)$ & $N ([R_{\rm OLR} - \Delta R, R_{\rm OLR}])$ & $N ([R_{\rm OLR}, R_{\rm OLR} + \Delta R])$ & ${\cal A}(R<R_{\rm OLR}) $ & ${\cal A}(R>R_{\rm OLR}) $ \\
        \hline
        0.0625 & 3,070,836 & 4,241,269 & +0.0075(14) & -0.0076(14)\\
        \hline
        0.0500 & 2,615,604 & 3,383,670 & +0.0068(14) & -0.0065(14)\\
        \hline
        0.0375 & 2,087,432 & 2,524,189 & +0.0058(15) & -0.0050(14)\\
        \hline
        0.0250 & 1,478,859 & 1,738,446 & +0.0050(16) & -0.0047(15)\\
        \hline
    \end{tabular}
    \caption{Star counts and background-corrected axial asymmetries for bins of varying width, $\Delta R$, probing just interior and exterior to the Outer Lindblad resonant radius, where the errors in the last digits are indicated in parentheses. As we focus in on the OLR, the magnitude of the asymmetry becomes slightly smaller, perhaps suggesting the magnitude of the first order radial correction, $|R_1|$, (see Eq. \ref{r1eq}) can be larger than a couple hundred parsecs. Also note that the radial bin external to the OLR has more stars due to the geometry of our 
    stellar sample \citep{GHY20, HGY20}.
    }
    \label{tab:DeltaRanalysis}
\end{table}

\section{Results} \label{sec:Results}

In this analysis, we have chosen the rotation curve of \citet{eilers2019circular} as it represents the only 
highly precise assessment of the 
Galaxy's rotation curve in the region of 
$5 \stackrel{<}{{}_{\sim}} R 
\stackrel{<}{{}_{\sim}} 25$ kpc.
As a check, we compute the Oort Constants, $A$ and $B$, using the $v_{\rm c}(R)$ parametrization in 
Eq.~\ref{vcparam} as given in \citet{eilers2019circular}
and find that $A = 14.95 \pm 0.43 \ \rm km \ s^{-1} \ kpc^{-1}$ and $B = -13.25 \pm 0.43 \ \rm km \ s^{-1} \ kpc^{-1}$, where we have combined the statistical and
$\pm 3\%$ systematic error in quadrature. 
These numbers are in very good agreement with the recent findings of \citet{li2019galactic} using {\it Gaia}
DR2 data within 500 pc of the Sun: 
 $A = 15.1 \pm 0.1 \ \rm km \ s^{-1} \ kpc^{-1}$ and $B = -13.4 \pm 0.1 \ \rm km \ s^{-1} \ kpc^{-1}$, though there is
 some tension in the determination of $B$ with respect to 
 the earlier results of \citet{binney2008GD} ($B = -12.4 \pm 0.6 \ \rm km \ s^{-1} \ kpc^{-1}$) and \citet{bovy2017galactic} ($B = -11.9 \pm 0.4  \ \rm km \ s^{-1} \ kpc^{-1}$). 

This rotation curve, along with a precise measurement of the Sun-GC distance \citep{abuter2019geometric} 
affords us the opportunity to use our OLR location 
determination to determine $\Omega_{\rm p}$ and also 
the location of the CR. 
Employing Eq.~\ref{PatternSpeed},
we have $\Omega_{\rm p} = 49.3 \pm 2.2 \ \rm km \ s^{-1} \ kpc^{-1}$.
By using our determined value of $R_{\rm OLR}$ and the leading order perturbation theory as per \citet{binney2008GD}, our pattern speed determination does not depend on any assumptions about the bar potential, 
other than its interpretation as a $m=2$ resonance. 
Moreover, the pattern speed we find falls within the literature average given by \citet{bland-hawthorne2016galaxy}, though it tends to be on the higher end as shown amongst a sample of other findings in Table~\ref{tab:patternspeedvals}.  
We recall, though, that  
as in the case of \citet{sanders2019pattern}, the pattern speed estimates can be biased low when including observations beyond the GC.

\begin{table}[]
    \centering
    \begin{tabular}{|c|c|c|c|}
        \hline
        Source & $\Omega_{\rm p}$ ($\rm km \ s^{-1} \ kpc^{-1}$) & Estimate of $R_{\rm CR}$ (kpc) & Estimate of $R_{\rm OLR}$ (kpc)\\
        \hline
        \citet{dehnen1999pattern} & $ 53 \pm 3$ & 4.44 & 7.34\\
        \hline
        \citet{sanders2019pattern} & $ 41 \pm 3$ & 5.69 & 9.32\\
        \hline
        \citet{sanders2019pattern} \footnote{Includes data from far side of the bar.} & $ 31 \pm 1$ & 7.43 & 12.01\\
        \hline
        \citet{hunt2018outer} ($m=4$)& $ \lesssim 1.35 \ \Omega_0 $\footnote{$\Omega_0 \approx 28 \rm \ km \ s^{-1} \ kpc^{-1}$ is the rotational frequency at the solar circle.} & $ > 6.15$ & $> 10.04$\\
        \hline
        \citet{portail2015mass} & $ 25 - 30 $ & $7.66 - 9.10$ & $12.37 - 14.54$\\
        \hline
        \citet{portail2016dynamical} & $ 39.0 \pm 3.5$ & 5.97 & 9.75\\
        \hline
        \citet{monari2017tracing} & $ > 1.8 \ \Omega_0 $ & $ < 4.66$ & $< 7.69$\\
        \hline
        \citet{chakrabarty2007phase} & $ 57.4^{+2.8}_{-3.3} $ & 4.11 & 6.81\\
        \hline
        \hline
        This work (without LMC correction) & $ 49.9 \pm 2.2 $ & $4.71 \pm 0.26$ &  $7.77 \pm 0.25$\\
        \hline
        This work (with LMC correction) & $ 49.3 \pm 2.2 $ & $4.76 \pm 0.27$ & $7.85 \pm 0.25$ \\
        \hline
        \hline
        \citet{bland-hawthorne2016galaxy} \footnote{Approximate literature range adopted in a review of galactic properties.} & $ 43 \pm 9 $ & 5.43 & 8.91\\
        \hline
    \end{tabular}
    \caption{The literature offers a wide array of pattern speed assessments.  
    The various assessments use differing assessments in 
    the Sun-GC distance and the local rotation curve, which could result in small changes. Also, 
    our CR and OLR estimates for each work use the rotation curve of \citet{eilers2019circular} and 
    the Sun-GC distance of \citet{abuter2019geometric}.}
    \label{tab:patternspeedvals}
\end{table}

Using this determined pattern speed in Eq.~\ref{CRcondition}, we estimate 
$R_{\rm CR} = (0.58 \pm 0.04) R_{0} = 4.76 \pm 0.27 \ \rm kpc$. 
Interestingly we determine that $R_{\rm OLR}/R_{\rm CR} \approx 1.7$
in agreement with the expectation 
of \citet{dehnen2000effect} if the bar is weak and the
rotation curve is flat. 
This is a useful consistency check as 
our CR determination is just compatible (within 1-$\sigma$) with the lower $R$ limit of the \citet{eilers2019circular} range of validity.  Additionally, this Corotation estimate is also just compatible within errors with the half-length of the bar, for which \citet{Wegg2015structure} find $\ell_{\rm bar} = 5.0 \pm 0.2$ kpc. We note that a weak bar should possess a CR at radii
beyond the half-length of the bar \citep{aguerri1998bar}. 
If the parameter $\delta$ is positive, 
reflective of a driving effect from a slowing
of the bar \citep{weinberg1993kinematic,chiba2019resonance}, then we can bring the picture into better agreement. The fluctuation of the bar's parameters suggested by \citet{hilmi2020fluctuations} could 
explain
the non-steady state effects we infer.

Given the diverse array of pattern speeds 
in the literature, as compiled in Table~\ref{tab:patternspeedvals}, it should perhaps come as no surprise that both the CR and the OLR have been argued to be near the solar circle.  As such, the wide spread in pattern speed assessments inevitably means that there are correspondingly large ranges for $R_{\rm CR}$ and $R_{\rm OLR}$.  Interestingly, though, a recent measurement of $R_{\rm OLR}$ by \citet{khoperskov2019hic} estimates the location of the OLR without assuming a pattern speed.  They find that the OLR is near $R = 9$ kpc, though they rely on models that draw random distributions of Gaia data that are very close to the mid-plane, for which the effects of reddening and extinction from dust would seem to be important.  
As an additional effect, the Milky Way's spiral arms break axial symmetry, but we have taken care to ensure that our sample is sufficiently out of plane so as to minimize any  confounding effects due to spiral structure \citep{GHY20}.

Finally, in addition to the pattern speed and the locations of the OLR and CR, we have found an unexpected, abrupt change in vertical structure near the OLR.  By computing the axial asymmetry for $z > 0$ and $z < 0$, henceforth the north (N) and south (S) respectively, we find 
as $R$ increases through the OLR, the asymmetry in the N
goes from left-heavy to right-heavy, with a smaller effect
of opposite sense in the S, 
as illustrated in Fig.~\ref{fig:verticalChange}.  Speculatively, this could be due to a vertical resonance with the bar, a bar tilted slightly out of plane, or perhaps stem from a North/South asymmetry in the bar itself, 
where we note that a North/South effect has been found 
in the Galactic center 
excess \citep{leane2020enigmatic}. 
Alternatively, local N/S differences have been noted in the solar neighborhood and have been attributed to the Sagittarius impact \citep{widrow2012galactoseismology,yanny2013stellar, ferguson2017milky}, so that the vertical effects seen near the OLR may come from a completely separate event.  Indeed, 
\citet{Carrillo2019kinematics} have suggested that the Sagittarius impact could have significantly perturbed the Galactic bar, or could have even been responsible for its genesis.
Detailed studies of the 
Galactic bar resonances in the presence of small vertical asymmetries in the bar or in the local disk, or subject to significant vertical perturbations could conceivably help explain this behavior.

We note that our assumption of an $m=2$ OLR resonance
can be
tested through additional observational studies. 
An $m=2$ OLR resonance 
 implies axially asymmetric structures at 
 $\phi=0, 180^\circ$, but the possibility of a 
 $m=4$ \citep{hunt2018outer} OLR 
 implies asymmetric structures at 
 $\phi=90, 270^\circ$ also, so that over 
 the longer term there is another observational 
 test \citep{hunt2018outer}. Yet this is not 
 the only possibility. Note that the existence of
 an $m=4$ resonance would imply that an $m=2$ resonance
 could appear (if it exists) at larger $R$ as well, so that 
 if our sign flip were interpreted as an $m=4$ resonance, 
 we would find $\Omega_{\rm p} \approx 39.3 \, \rm km \, s^{-1} \, kpc^{-1}$ and 
 a $m=2$ resonance at $R_{\rm OLR} \approx 11.6$ kpc.  This alternative possibility meshes well with the findings of \citet{portail2016dynamical} and could be explored in future data releases.

\section{Summary} \label{sec:summary}

We have shown 
that axial symmetry breaking orbital alignments are detectable at very small levels and 
that our analysis of this effect is 
consistent with leading order perturbation theory 
that models the Galactic bar as a weakly non-axially symmetric effect.  
Through this approach, we avoid the need to assume a form for the galaxy's potential, apart from the 
assumption of a $m=2$ potential,  
and we only rely on the quadrant in which the bar points in order to interpret 
the sign flip we observe in the asymmetry. 
We have found that the OLR is situated at $R_{\rm OLR} = 7.85 \pm 0.25 \ \rm kpc$, which implies the pattern speed of the bar is $\Omega_{\rm p} = 49.3 \pm 2.2 \rm \ km \ s^{-1} \ kpc^{-1}$, and thus the radius of Corotation is $R_{\rm CR} = 4.76 \pm 0.27 \rm \ kpc $.  
Additionally, we find evidence for a change in the vertical structure of the disk near the OLR, but we cannot resolve if this effect is due to a possibly tilted or asymmetric bar, or if the effect is local in nature, possibly due to the Sagittarius impact. 
Our approach is entirely novel, but 
our estimates for the pattern speed of the bar are very much consistent with the upward revision of the $\Omega_{\rm p}$ of \citet{sanders2019pattern} and
\citet{bovy2019life} as suggested by the work of \citet{hilmi2020fluctuations},
and our inferred resonance locations for the CR and the OLR are in remarkable agreement with the picture of \citet{dehnen1999pattern}, even if
our assessments are much more precise. 
Thus we believe that our results are in support of a 
Galactic bar that is both weak and fast.

\acknowledgements{

A.H. thanks the Universities Research Association 
for travel to Fermilab, the GAANN Fellowship for support, and Isaac Shlosman for discussion of the resonances of the Galactic bar.
S.G. and A.H. acknowledge partial support from the U.S.
Department of Energy under contract DE-FG02-96ER40989, 
and S.G. also acknowledges the University Research Professor fund of the University of Kentucky for partial support. 
The authors also thank the anonymous referee for helpful comments.

This document was prepared in part using the resources of Fermi National Accelerator Laboratory (Fermilab), a U.S. Department of Energy, Office of Science, HEP User Facility. Fermilab is managed by the Fermi Research Alliance, LLC (FRA), acting under Contract No. DE-AC02-07CH11359. 

This work has made use of data from the European Space Agency (ESA) mission
{\it Gaia} (\url{https://www.cosmos.esa.int/gaia}), processed by the {\it Gaia}
Data Processing and Analysis Consortium (DPAC,
\url{https://www.cosmos.esa.int/web/gaia/dpac/consortium}). Funding for the DPAC
has been provided by national institutions, in particular the institutions
participating in the {\it Gaia} Multilateral Agreement.
}

\bibliography{mybib.bib}
\end{document}